# Compressed Air Energy Storage-Part I: An Accurate Bi-linear Cavern Model

Junpeng Zhan, *Member, IEEE,* Osama Aslam Ansari, *Student Member, IEEE,*
and C. Y. Chung, *Fellow, IEEE*

***Abstract*—Compressed air energy storage (CAES) is suitable for large-scale energy storage and can help to increase the penetration of wind power in power systems. A CAES plant consists of compressors, expanders, caverns, and a motor/generator set. Currently used cavern models for CAES are either accurate but highly non-linear or linear but inaccurate. Highly non-linear cavern models cannot be directly utilized in power system optimization problems. In this regard, an accurate bi-linear cavern model for CAES is proposed in this first paper of a two-part series. The charging and discharging processes in a cavern are divided into several virtual states and then the first law of thermodynamics and ideal gas law are used to derive a cavern model, i.e., model for the variation of temperature and pressure in these processes. Thereafter, the heat transfer between the air in the cavern and the cavern wall is considered and integrated into the cavern model. By subsequently eliminating several negligible terms, the cavern model reduces to a bi-linear (linear) model for CAES with multiple (single) time steps. The accuracy of the proposed cavern model is verified via comparison with an accurate non-linear model.**

***Index Terms*—Bi-linear cavern model; compressed air energy storage (CAES); heat transfer; ideal gas law; thermodynamics.**

## NOMENCLATURE

|  |  |
|---|---|
| a,ht | Both adiabatic process and heat transfer are considered (superscript) |
| ht | Heat transfer (superscript) |
| $a_i$ | Parameters, $i = 2,3,4$ |
| $c_i$ | Parameters, $i = 0,1$, representing the left-hand side of (3) and (10), respectively |
| $c_v$ | Constant volume specific heat (J/(kg K)) |
| $h_c$ | Heat transfer coefficient (W/(m$^2$ K)) |
| $k$ | A constant equal to 1.4 |
| $m_s$ | Mass of air in the cavern (kg) |
| $m_o$ | Mass of air in virtual container 2 as shown in Fig. 3 (kg) |
| $m_{in}$ | Mass of air charged into the cavern (kg) |
| $\dot{m}_{in}$ | Mass flow rate charged into a cavern (kg/s) |
| $\dot{m}_{out}$ | Mass flow rate discharged out of a cavern (kg/s) |
| $p_s$ | Pressure of the air in the cavern (bar) |
| $p_{si}$ | Pressure of the air in the cavern after the charging, discharging, and idle processes for $i$ =2, 3, and 4, respectively (bar) |
| $p_{\text{in}}$ | Pressure of the air charged into the cavern (bar) |
| $p_{inj}$ | Pressures in virtual states as shown in Fig. 2, $j = 1,2$ (bar) |
| $A_c$ | Surface area of the cavern wall (m$^2$) |
| $Q$ | Total internal energy (J) |
| $R$ | Specific air constant (J/(kg K)) |
| $T_s$ | Temperature of the air in the cavern (K) |
| $T_{si}$ | Temperature of the air in the cavern after the charging, discharging, and idle processes for $i$ =2, 3, and 4, respectively (K) |
| $T_{RW}$ | Temperature of the cavern wall (K) |
| $T_{\text{in}}$ | Temperature of the air charged into the cavern (K) |
| $T_{inj}$ | Temperatures in virtual states as shown in Fig. 2, $j =$ 1,2 (K) |
| $V_{in1}, V_o$ | Volumes of virtual containers as shown in Figs. 2 and 3 (m$^3$) |
| $V_s$ | Volume of a cavern (m$^3$) |
| $W$ | Work (J) |
| $\rho_{av}$ | Average air density in a cavern (kg/m$^3$) |
| $\Delta t$ | Time interval (s) |
| $\Delta U$ | Change in internal energy (J) |

The work was supported in part by the Natural Sciences and Engineering Research Council (NSERC) of Canada and the Saskatchewan Power Corporation (SaskPower).

The authors were with the Department of Electrical and Computer Engineering, University of Saskatchewan, Saskatoon, SK S7N 5A9, Canada (e-mail: zhanjunpeng@gmail.com, oa.ansari@usask.ca, c.y.chung@usask.ca).

## I. INTRODUCTION

ENERGY storage technologies are valuable to power systems, especially considering the penetration of renewable generation is growing rapidly, e.g., the wind power share of global electricity demand will increase from 4% in 2015 to 25-28% in 2050 [1]. Energy storage can provide different kinds of services [2], e.g., electric energy time-shift, electric supply capacity, regulation, power reliability, etc. The current global installed electricity storage capacity is about 141 GW and an estimated 310 GW of additional capacity would be needed in the United States, Europe, China, and India [3] to support the massive increase of renewable generation in the future. There are currently two kinds of large-scale energy storage, i.e., pumped-hydro storage and compressed air energy storage (CAES), that can be installed at the grid scale.

CAES is a high power and energy storage technology and has relatively low capital, operational, and maintenance costs [4]. The power rating of a large-scale CAES plant can reach 300 or even 1000 MW and the rated energy capacity can reach 1000 or even 2860 MWh [4]. Currently, there are two commercialized CAES plants. The world's first CAES plant was installed in Huntorf, Germany in 1978. A schematic of the Huntorf CAES plant is shown in Fig. 1 [5]. A CAES plant is comprised of compressors, turbines, a motor/generator set, and large repositories, e.g., underground salt caverns. CAES uses off-peak electricity (up to 60 MW for the Huntorf CAES plant) to compress the air to high pressure and store it in a large repository. CAES generates electricity (up to 290 MW for the Huntorf CAES plant) by releasing the stored compressed air, which is combusted with fuel to drive the turbines. The second commercialized CAES plant is the McIntosh plant in McIntosh, Alabama, U.S. [6]. This plant, which became operational in 1991, can produce an output of 110 MW electricity for up to 26 hours. The plant efficiencies of the Huntorf plant and McIntosh plant are ~42% and ~54%, respectively [7]. The round-trip efficiency of the CAES is ~80% [8].

According to the Electric Power Research Institute (EPRI), about 75% of the U.S. has geologic sites suitable for CAES [7], [8]. Northern Europe is also replete with suitable salt deposits. For example, nearly 500 salt caverns are currently being used for natural gas storage. Therefore, it is feasible to install CAES in many different locations. A number of CAES plants are being constructed as a result of increasing renewable energy utilization and several advantages offered by CAES. For example, a 160 MW CAES plant near Saskatchewan-Alberta border in Canada [9] is expected to be completed in a few years and combined with the interconnection between the Saskatchewan and Alberta power grids. From 2009 to 2013, Pacific Gas & Electric received US$50 million in funding for a demonstration project to validate the design, performance, and reliability of a 300 MW CAES plant in Kern County, California [10]. Several further examples are provided in [10].

Thermodynamic properties such as variations of temperature and pressure in the caverns of a CAES plant are important factors that affect the overall plant operation and performance [11]-[13]. Two kinds of cavern models for CAES are currently described in the literature.

The first consists of accurate but highly nonlinear models [11]-[13]. In [11], complex and simplified real gas models are developed for an adiabatic cavern for CAES, both of which adequately represent the thermodynamic properties of the air. Reference [12] developed an accurate dynamic simulation model for a CAES cavern that incorporates an accurate heat transfer model. In [12], heat transfer is shown to play an important role in the thermodynamic behavior of the cavern and therefore the proposed model can accurately simulate the actual cavern behavior. In [13], a simplified and unified analytical solution considering heat transfer is proposed for temperature and pressure variations in CAES caverns. The model proposed in [13] is validated using real data from the Huntorf plant trial tests and the results calculated from the models in [11] and [12], demonstrating that the proposed solution is capable of adequately calculating the thermodynamic behavior of CAES caverns. All three models in [11]-[13] are accurate but highly non-linear, and therefore cannot be used in large-scale power system optimization problems.

The second kind of cavern model assumes that the air temperature in the cavern is constant. This kind of model has been adopted in different power system operation problems, e.g., transmission congestion relief [14], bidding and offering strategy [15], and unit commitment [16]. The constant temperature model is linear but inaccurate, which can result in non-optimal or even infeasible solutions.

In this regard, a novel bi-linear cavern model based on the ideal gas law and the first law of thermodynamics is proposed in this paper (the first in a two-part series), where the heat transfer between the air in the cavern and the cavern wall is considered. The advantages of the bi-linear model over the existing two types of models mentioned above are two-fold: 1) it is accurate, as will be verified in this paper, and 2) it can be integrated into large-scale power system optimization problems, as will be demonstrated in the second paper of this two-part series. The main contribution of this paper is the proposed accurate bi-linear cavern model of the CAES.

The rest of this paper is organized as follows. Section II details the deduction of the accurate bi-linear cavern model. Section III verifies the effectiveness of the proposed cavern model. Section IV presents the conclusions drawn from the results.

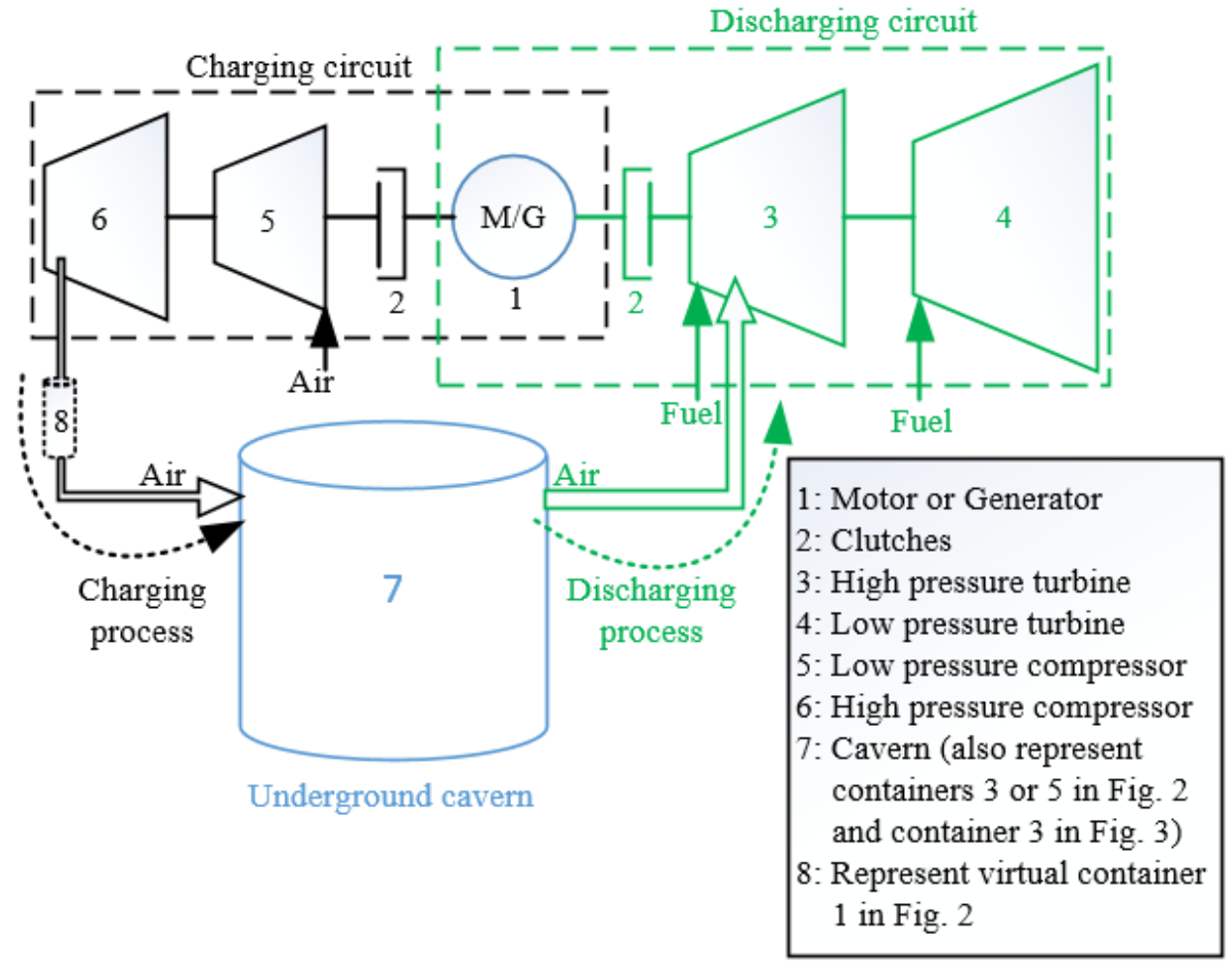

Fig. 1. Schematic of the Huntorf CAES plant.

## II. Accurate Bi-linear Cavern Model for CAES

In this paper, constant-volume caverns of CAES are considered, as they are used in the existing CAES plants.

### A. *Charging Process*

In the charging process, a certain amount of air is injected by compressors into a cavern, as shown in Fig. 1. To facilitate the model deduction given in Sections II-A1 to II-A3, the charging process is divided into four states and the air is assumed to be stored in the (virtual) containers as shown in Fig. 2, where the five containers are indexed by the numbers in heptagons. The characteristics of the air in each container, including the pressure, volume, temperature, and mass, are shown in each container in Fig. 2. The values of the underlined notations, i.e., the pressure and temperature in containers 2, 4, and 5, are not known while the values of the other notations are known.

Containers 1, 2, and 4 are virtual while containers 3 and 5 represent the cavern before and after the air from the compressors is injected into it, respectively.

It is assumed that the air coming out of the compressors is stored in a virtual container, i.e., cylinder 8 in Fig. 1 and container 1 in Fig. 2. The conditions of container 1 represent the thermodynamic properties of air at the outlet of the compressors. First, let the air in virtual container 1 go into virtual container 2. The volume of container 2 is set such that the ratio of the volume of containers 2 to 3 is equal to the ratio of the mass of air in containers 2 to 3, i.e.,

$$V_{in1}/V_s = \dot{m}_{in}\Delta t/m_s \tag{1}$$

where $V_{in1}$ and $V_s$ represent the volumes of containers 2 and 3, $m_s$ represents the mass of air in container 3, and $\dot{m}_{in}$ represents the flow rate of mass charged into the cavern, which is assumed to be constant during a period of time, $\Delta t$. The mass coming out of the compressor during that period of time, denoted as $m_{in}$, can be expressed as $m_{in} = \dot{m}_{in}\Delta t$. The mass of air injected into the cavern is assumed to be equal to $m_{in}$, i.e., there is no air leakage. $m_{in}$ is also the mass of air in both containers 1 and 2.

Then, let the air in both containers 2 and 3 go into container 4. The volume of (mass of air in) container 4 is set to the sum of the volumes of (mass of air in) containers 2 and 3, i.e., container 4 can be seen as a combination of containers 2 and 3. Note that the purpose of using virtual containers 2 and 4 is to let the work be 0 during the process of merging the air in containers 2 and 3 into container 4.

Last, let the air in container 4 go into container 5, which is an adiabatic compressing process as the mass of air does not change and the volume decreases. The rest of this subsection details the deduction of the model for the charging process.

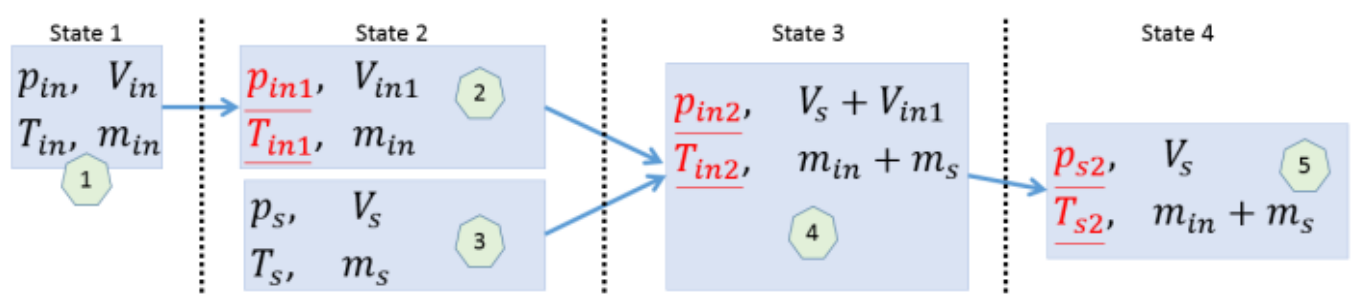

Fig. 2. Four states and five containers used for model deduction in the charging process.

*1) State 1→State 2*

The transfer of air from containers 1 to 2 is an adiabatic process and the mass of air does not change. According to the ideal gas law for the air in container 2, one can obtain

$$p_{in1}V_{in1} = \dot{m}_{in}\Delta t R T_{in1}. \tag{2}$$

According to the temperature-pressure relation for an adiabatic process, $T^{\frac{k}{k-1}}/p$ is constant from containers 1 to 2 [17] and one can obtain

$$\frac{(T_{in})^{\frac{k}{k-1}}}{p_{in}} = \frac{(T_{in1})^{\frac{k}{k-1}}}{p_{in1}}. \tag{3}$$

Let $c_0$ represent the left-hand side of (3), i.e., $c_0 = T_{in}^{\frac{k}{k-1}}/p_{in}$. $V_{in1}$ can be determined from (1), i.e., $V_{in1} = V_s\dot{m}_{in}\Delta t/m_s$. This leaves only two unknown variables in (2) and (3), i.e., $p_{in1}$ and $T_{in1}$. Therefore, $p_{in1}$ and $T_{in1}$ can be obtained from (2) and (3):

$$p_{in1} = (c_0)^{k-1}R^k m_s^k/V_s^k \tag{4}$$

$$T_{in1} = (c_0 R m_s/V_s)^{k-1}. \tag{5}$$

*2) State 2→State 3*

Now we consider the process of the air in both containers 2 and 3 going into container 4. The volume of container 4 is equal to the sum of the volumes of containers 2 and 3. In this process, the work is zero and the total internal energy does not change. The change of the internal energy [18] in containers 2 and 3 is $m_{in}c_v(T_{in2} - T_{in1})$ and $m_s c_v(T_{in2} - T_s)$, respectively. According to the first law of thermodynamics, i.e., $Q = \Delta U + W$, one can obtain

$$m_{in}c_v(T_{in2} - T_{in1}) + m_s c_v(T_{in2} - T_s) = 0. \tag{6}$$

Note that $T_{in1}$ can be obtained from (5). $T_{in2}$ is then the only unknown variable in (6) and can be expressed as

$$T_{in2} = (m_{in}T_{in1} + m_s T_s)/(m_{in} + m_s). \tag{7}$$

According to the ideal gas law for the air in container 4 in Fig. 2, one can obtain

$$p_{in2}(V_s + V_{in1}) = (m_{in} + m_s)RT_{in2} \tag{8}$$

Then, by substituting (7) into (8), one can obtain

$$p_{in2} = \frac{(m_{in}+m_s)RT_{in2}}{V_s+V_{in1}} = \frac{(m_{in}T_{in1}+m_s T_s)R}{V_s+V_{in1}} \tag{9}$$

Therefore, $p_{in2}$ and $T_{in2}$ in container 4 are obtained.

*3) State 3→State 4*

From containers 4 to 5, the mass of air does not change and the volume reduces from $(V_s + V_{in1})$ to $V_s$. This is an adiabatic compression/heating process. Thus, $T^{\frac{k}{k-1}}/p$ is constant from containers 4 to 5 [17], i.e.,

$$\frac{(T_{in2})^{\frac{k}{k-1}}}{p_{in2}} = \frac{(T_{s2})^{\frac{k}{k-1}}}{p_{s2}} \tag{10}$$

According to the ideal gas law for the air in container 5 in Fig. 2, one has

$$p_{s2}V_s = (\dot{m}_{in}\Delta t + m_s)RT_{s2} \tag{11}$$

Let $c_1$ represent the left-hand side of (10), i.e., $c_1 = (T_{in2})^{\frac{k}{k-1}}/p_{in2}$. Only two variables are unknown in (10) and (11), i.e., $p_{s2}$ and $T_{s2}$. Therefore, $p_{s2}$ and $T_{s2}$ can be obtained from (10) and (11):

$$p_{s2} = (\dot{m}_{in}\Delta t + m_s)^k R^k (c_1)^{k-1}/V_s^k \tag{12}$$

$$T_{s2} = (c_1 p_{s2})^{1-\frac{1}{k}}. \tag{13}$$

Now $p_{s2}$ and $T_{s2}$ for container 5 have been obtained. By substituting $c_1$ ($c_0$, (5), (7), and (9) are needed to calculate $c_1$) into (12) and (13), these two equations can be reformed as:

$$p_{s2} = p_s\left(1 + \frac{\dot{m}_{in}\Delta t}{m_s}\right)^{k-1} + a_2(\dot{m}_{in}\Delta t + m_s)^{k-1}\dot{m}_{in}\Delta t \tag{14}$$

$$T_{s2} = T_s\left(1 + \frac{\dot{m}_{in}\Delta t}{m_s}\right)^{k-2} + a_3(\dot{m}_{in}\Delta t + m_s)^{k-2}\dot{m}_{in}\Delta t \tag{15}$$

where $a_2 = \frac{R^k T_{in}^k}{V_s^k p_{in}^{k-1}}$ and $a_3 = \frac{R^{k-1}T_{in}^k}{V_s^{k-1}p_{in}^{k-1}}$.

Equations (14) and (15) show that $p_{s2}$ and $T_{s2}$ are nonlinear functions of $\dot{m}_{in}$, which are linearized as follows.

According to Newton's generalized binomial theorem [19], one has

$$(1 + x)^r = \sum_{j=0}^{\infty}\binom{r}{j}x^j = 1 + rx + \frac{r(r-1)}{2!}x^2 + \cdots \tag{16}$$

where $\binom{r}{j} = \frac{r(r-1)\cdots(r-j+1)}{j!}$, $r$ can be any real number, and $j$ is an integer. That is, $\left(1 + \frac{\dot{m}_{in}\Delta t}{m_s}\right)^{k-1}$ in (14) can be expressed as $\left(1 + (k-1)\frac{\dot{m}_{in}\Delta t}{m_s} + \frac{(k-1)(k-2)}{2!}\left(\frac{\dot{m}_{in}\Delta t}{m_s}\right)^2 + \cdots\right)$. Considering that $\dot{m}_{in}\Delta t$ is much smaller than $m_s$, the second and higher orders of $(\dot{m}_{in}\Delta t/m_s)$ can be ignored. Then, (14) can be reformed as

$$p_{s2} = p_s\left(1 + (k-1)\frac{\dot{m}_{in}\Delta t}{m_s}\right) + a_2(m_s)^{k-1}\dot{m}_{in}\Delta t \tag{17}$$

Note that the second term in (14) is replaced by $a_2(m_s)^{k-1}\dot{m}_{in}\Delta t$, which has negligible error as $a_2$ is much smaller than $p_s$ (e.g., $p_s$ =46~66× $10^5$ and $a_2$ =1.04× $10^{-3}$ for the Huntorf CAES plant) and $\dot{m}_{in}\Delta t$ is much smaller than $m_s$. Similarly, (15) can be reformed as

$$T_{s2} = T_s\left(1 + (k-2)\frac{\dot{m}_{in}\Delta t}{m_s}\right) + a_3(m_s)^{k-2}\dot{m}_{in}\Delta t \tag{18}$$

When (17) and (18) are used in a one-step optimization problem, $p_{s2}$ is a linear function of $\dot{m}_{in}$ in (17) and $T_{s2}$ is a linear function of $\dot{m}_{in}$ in (18) as $p_s$ and $T_s$ have a known initial status. When used in a multi-step optimization problem, (17) and (18) are bi-linear equations as $p_s$ and $T_s$ become decision variables.

### *B. Discharging Process*

$p_s,\ V_s - V_o$
$T_s,\ m_s - m_o$ (1)

$p_s,\ V_o$
$T_s,\ m_o$ (2)

$p_{s3},\ V_s$
$T_{s3},\ m_s - m_o$ (3)

Fig. 3. Three containers used for model deduction in the discharging process.

In the discharging process, the air is discharged from a cavern, as shown in Fig. 1. To facilitate the model deduction given in the rest of this subsection, the cavern before discharging is divided into two containers, i.e., containers 1 and 2, while the cavern after discharging (cylinder 7 in Fig. 1) is represented as container 3, as shown in Fig. 3. The three containers in Fig. 3 are indexed by the numbers in heptagons. The characteristics of the air in each container, including the pressure, volume, temperature, and mass, are shown in each container in Fig. 3. The values of the underlined notations, i.e., the pressure and temperature in container 3, are not known while the values of the other notations are known.

The discharging process can be divided into two virtual steps. First, the air in container 2 is taken out of the cavern. Then, the air in container 1 expands to the whole cavern, i.e., air goes from containers 1 to 3. The deduction of the model for the second step is given in the rest of this subsection.

The volume of container 2 is set such that the ratio of the volume of containers 2 to 1 is equal to the ratio of the mass of air in containers 2 to 1, i.e.

$$V_o/(V_s - V_o) = m_o/(m_s - m_o). \tag{19}$$

Note that the purpose of using virtual container 2 is to let the temperature and pressure of the air in container 1 be the same as the air in the cavern before discharging and to let air going from containers 1 to 3, i.e., step 2, be an adiabatic expansion process.

Let $\dot{m}_{out}$ represent the flow rate of mass discharged from the cavern, which is assumed to be constant during a period of time, $\Delta t$. Then, the mass of air discharged from the cavern, denoted as $m_o$, during that period of time can be expressed as $m_o = \dot{m}_{out}\Delta t$.

Air expansion from containers 1 to 3 is an adiabatic process. Then, $T^{\frac{k}{k-1}}/p$ is constant from containers 1 to 3:

$$\frac{(T_s)^{\frac{k}{k-1}}}{p_s} = \frac{(T_{s3})^{\frac{k}{k-1}}}{p_{s3}}. \tag{20}$$

According to the ideal gas law for the air in container 3 in Fig. 3, one has

$$p_{s3}V_s = (m_s - m_o)RT_{s3}. \tag{21}$$

There are only two variables unknown in (20) and (21), i.e., $p_{s3}$ and $T_{s3}$. Therefore, $p_{s3}$ and $T_{s3}$ can be obtained from (20) and (21):

$$p_{s3} = (1 - \dot{m}_{out}\Delta t/m_s)^k p_s \tag{22}$$

$$T_{s3} = (1 - \dot{m}_{out}\Delta t/m_s)^{k-1} T_s. \tag{23}$$

According to the Newton's generalized binomial theorem, when $\dot{m}_{out}\Delta t$ is much smaller than $m_s$, (22) and (23) can be respectively reformed as

$$p_{s3} = (1 - k\dot{m}_{out}\Delta t/m_s)p_s \tag{24}$$

$$T_{s3} = (1 - (k-1)\dot{m}_{out}\Delta t/m_s)T_s \tag{25}$$

Similar to (17) and (18), (24) and (25) are linear (bi-linear) equations when used in a one-step (multi-step) optimization problem.

### *C. Charging Process Considering Heat Transfer*

In Sections II-A and II-B, the heat transfer between the air and the cavern wall is not considered. However, the heat transfer plays an important role in the variation of the air temperature/pressure in the cavern [12]. Therefore, the heat transfer is considered in this and the following two subsections. In this subsection, the temperature as a function of time is first deduced. The pressure as a function of time is then obtained via the ideal gas law. Last, the temperature/pressure as a function of time is linearized to obtain a bi-linear model.

According to [13], the air density ($\rho_{av}$) in the cavern and the cavern wall temperature ($T_{RW}$) can be assumed to be constant and the heat transfer between the air and the cavern wall can be modelled as

$$\frac{dT}{dt} = \frac{h_c A_c}{V_s \rho_{av} c_v}(T_{RW} - T) \tag{26}$$

$$T(t) = \int \frac{h_c A_c}{V_s \rho_{av} c_v}(T_{RW} - T)dt \tag{27}$$

Equation (18) can be written as

$$T_{s2}(t) = T_s\left(1 + (k-2)\frac{\dot{m}_{in}\, t}{m_s}\right) + a_3(m_s)^{k-2}\dot{m}_{in}\, t. \tag{28}$$

By substituting (28) into (27), i.e., replacing $T$ on the right-hand side of (27) by the right-hand side of (28), one can obtain

$$T^{ht}_{s2}(t) = \int \frac{h_c A_c}{V_s \rho_{av} c_v}\left(T_{RW} - T_s\left(1 + (k-2)\frac{\dot{m}_{in}\, t}{m_s}\right) - a_3(m_s)^{k-2}\dot{m}_{in}\, t\right)dt \tag{29}$$

where superscript 'ht' represents 'heat transfer'.

By solving the integral equation (29), one can obtain

$$T^{ht}_{s2}(t) = \frac{h_c A_c}{V_s \rho_{av} c_v}\left(T_{RW}\, t - T_s\left(t + (k-2)\frac{\dot{m}_{in}\, t^2}{2m_s}\right) - a_3(m_s)^{k-2}\dot{m}_{in}\, t^2/2\right) \tag{30}$$

Adding (28) and (30) together gives

$$T^{\text{a,ht}}_{s2}(t) = T_s\left(1 + (k-2)\frac{\dot{m}_{in}\, t}{m_s}\right) + a_3(m_s)^{k-2}\dot{m}_{in}\, t + \frac{h_c A_c}{V_s \rho_{av} c_v}\left(T_{RW}\, t - T_s\left(t + (k-2)\frac{\dot{m}_{in}\, t^2}{2m_s}\right) - a_3(m_s)^{k-2}\dot{m}_{in}\frac{t^2}{2}\right) \tag{31}$$

where superscript 'a,ht' indicates that both the adiabatic process and heat transfer are considered.

According to [19], i.e., $f(x) = f(x_0) + f'(x_0) \times (x - x_0)$, one can linearize $(m_s)^{k-1}$ and $(m_s)^{k-2}$ at $m_{av0}$ as

$$(m_s)^{k-1} = (m_{av0})^{k-1} + (k-1)(m_{av0})^{k-2}(m_s - m_{av0}) \tag{32}$$

$$(m_s)^{k-2} = (m_{av0})^{k-2} + (k-2)(m_{av0})^{k-3}(m_s - m_{av0}) \tag{33}$$

where $m_{av0}$ is a fixed value, i.e., $m_{av0} = \rho_{av}V_s$.

Then, by using (32) and (33), (31) can be reformed as

$$m_s T^{\text{a,ht}}_{s2}(t) = T_s(m_s + (k-2)\dot{m}_{in}\, t) + a_3\dot{m}_{in}\, t((m_{av0})^{k-1} + (k-1)(m_{av0})^{k-2}(m_s - m_{av0})) + \frac{h_c A_c}{c_v}(T_{RW}\, t - T_s(t + 0.5(k-2)\dot{m}_{in}\, t^2/m_{av0}) - 0.5a_3\dot{m}_{in}t^2((m_{av0})^{k-2} + (k-2)(m_{av0})^{k-3}(m_s - m_{av0}))) \tag{34}$$

Equation (34) represents the change of the temperature during the charging process as a function of time $t$ and charging mass flow rate $\dot{m}_{in}$, where both the adiabatic process and the heat transfer process are considered.

According to the ideal gas law, one can obtain

$$p^{\text{a,ht}}_{s2}(t) = (m_s + \dot{m}_{out}t)RT^{\text{a,ht}}_{s2}(t)/V_s \tag{35}$$

which can be expanded to (36) by substituting (31) therein.

$$p^{\text{a,ht}}_{s2}(t) = \frac{(m_s + \dot{m}_{in}t)RT_s}{V_s}\left(1 + (k-2)\frac{\dot{m}_{in}\, t}{m_s}\right)$$

$$+\frac{(m_s+\dot{m}_{in}t)R}{V_s}a_3(m_s)^{k-2}\dot{m}_{in}\,t$$
$$+\frac{(m_s+\dot{m}_{in}t)R}{V_s}\frac{h_cA_c}{V_s\rho_{av}c_v}\left(T_{RW}\,t-T_st-T_s(k-2)\frac{\dot{m}_{in}\,t^2}{2m_s}\right)$$
$$-\frac{(m_s+\dot{m}_{in}t)R}{V_s}\frac{h_cA_c}{V_s\rho_{av}c_v}\left(a_3(m_s)^{k-2}\dot{m}_{in}\frac{t^2}{2}\right) \quad (36)$$

The four terms in (36) are each on a separate line. The second term of (36) can be replaced by $\frac{m_sR}{V_s}a_3(m_s)^{k-2}\dot{m}_{in}\,t$ because $Ra_3/V_s$ is small and $\dot{m}_{in}t$ is much smaller than $m_s$. The last term of (36) can be ignored because both $\frac{h_cA_c}{V_s\rho_{av}c_v}$ and $Ra_3/V_s$ are small.

According to [19], i.e., $f(x)=f(x_0)+f'(x_0)\times(x-x_0)$, one can linearize $m_s^k$ at $m_{av0}$:

$$m_s^k=m_{av0}^k+km_{av0}^{k-1}(m_s-m_{av0}) \quad (37)$$

Now, by using (37), (36) can be reformed as

$$m_sp_{s2}^{\mathrm{a,ht}}(t)=p_s(m_s+(k-1)\dot{m}_{in}t)+a_2\dot{m}_{in}\,t\left(m_{av0}^k+km_{av0}^{k-1}(m_s-m_{av0})\right)+\frac{h_cA_c}{c_v}\big((m_s+0.5\dot{m}_{in}t)T_{RW}tR/V_s-p_st-0.5(k-1)\dot{m}_{in}T_st^2R/V_s\big) \quad (38)$$

## D. Discharging Process Considering Heat Transfer

In this subsection, the temperature as a function of time is first deduced. The pressure as a function of time is then obtained via the ideal gas law. Last, the temperature/pressure as a function of time is linearized to obtain a bi-linear model.

Equation (25) can be written as

$$T(t)=T_s-(k-1)T_s\frac{\dot{m}_{out}}{m_s}t \quad (39)$$

By substituting (39) into (27), i.e., replacing $T$ on the right-hand side of (27) by the right-hand side of (39), one can obtain

$$T_{s3}^{ht}(t)=\int\frac{h_cA_c}{V_s\rho_{av}c_v}\left(T_{RW}-T_s+(k-1)T_s\frac{\dot{m}_{out}}{m_s}t\right)dt \quad (40)$$

By solving the integral equation (40), one can obtain

$$T_{s3}^{ht}(t)=\frac{h_cA_c}{V_s\rho_{av}c_v}\left((T_{RW}-T_s)t+(k-1)T_s\frac{\dot{m}_{out}}{2m_s}t^2\right) \quad (41)$$

Adding (39) and (41) together gives

$$T_{s3}^{\mathrm{a,ht}}(t)=T_s-(k-1)T_s\frac{\dot{m}_{out}}{m_s}t+\frac{h_cA_c}{V_s\rho_{av}c_v}\left((T_{RW}-T_s)t+(k-1)T_s\frac{\dot{m}_{out}}{2m_s}t^2\right) \quad (42)$$

Equation (42) represents the change in temperature during the discharging process as a function of time $t$ and discharging mass flow rate $\dot{m}_{out}$, where both the adiabatic process and the heat transfer process are considered. Considering that $\frac{h_cA_c}{V_s\rho_{av}c_v}$ is very small (around $1\times10^{-4}$), $\frac{h_cA_cm_s}{V_s\rho_{av}c_v}$ can be replaced by $\frac{h_cA_c}{c_v}$ and therefore (42) can be reformed as

$$m_sT_{s3}^{\mathrm{a,ht}}(t)=m_sT_s-(k-1)T_s\dot{m}_{out}t+\frac{h_cA_c}{c_v}(T_{RW}-T_s)t+\frac{h_cA_c}{2m_{av0}c_v}(k-1)T_s\dot{m}_{out}t^2 \quad (43)$$

Note that there are four variables in (43), i.e., $m_s$, $\dot{m}_{out}$, $T_{s3}^{\mathrm{a,ht}}(t)$, and $T_s$.

According to the ideal gas law, one can obtain

$$p_{s3}^{\mathrm{a,ht}}(t)=(m_s-\dot{m}_{out}t)RT_{s3}^{\mathrm{a,ht}}(t)/V_s \quad (44)$$

which can be expanded as follows by substituting (42) therein:

$$p_{s3}^{\mathrm{a,ht}}(t)=\frac{(m_s-\dot{m}_{out}t)RT_s}{V_s}-\frac{(m_s-\dot{m}_{out}t)R}{V_s}(k-1)T_s\frac{\dot{m}_{out}}{m_s}t+\frac{h_cA_c}{V_s\rho_{av}c_v}\left(\frac{(m_s-\dot{m}_{out}t)R}{V_s}(T_{RW}-T_s)t+\frac{(m_s-\dot{m}_{out}t)R}{V_s}(k-1)T_s\frac{\dot{m}_{out}}{2m_s}t^2\right) \quad (45)$$

Note that $(m_s-\dot{m}_{out}t)\frac{\dot{m}_{out}}{m_s}$ in the second term of (45) can be replaced by $m_s\frac{\dot{m}_{out}}{m_s}$ because $\dot{m}_{out}t$ is much smaller than $m_s$. Equation (45) can be reformed as

$$m_sp_{s3}^{\mathrm{a,ht}}(t)=(m_s-k\dot{m}_{out}t)p_s+\frac{h_cA_cR}{c_vV_s}\big((m_s-0.5\dot{m}_{out}t)(T_{RW}-T_s)t+0.5(k-1)T_s\dot{m}_{out}t^2\big) \quad (46)$$

Comparing (46) with (24), we know that the first term in (46) represents the adiabatic process inside the cavern and the second term is associated with the heat transfer between the air in the cavern and the cavern wall. Note that there are five variables in (46), i.e., $m_s$, $\dot{m}_{out}$, $p_{s3}^{\mathrm{a,ht}}(t)$, $p_s$, and $T_s$.

## E. Idle Process Considering Heat Transfer

When in the idle process, i.e., neither charging nor discharging occurs, heat transfer occurs between the air and the cavern wall if there is a temperature difference between them. By solving the integral equation (27), the change of temperature in the cavern in the idle process can be expressed as

$$T_{s4}^{\mathrm{ht}}(t)=(T_s-T_{RW})e^{-\frac{h_cA_c}{V_s\rho_{av}c_v}t}+T_{RW} \quad (47)$$

where $T_s$ is the initial temperature of the air in the cavern in the idle process.

According to the ideal gas law, one can obtain

$$p_{s4}^{\mathrm{ht}}(t)=m_sRT_{s4}^{\mathrm{ht}}(t)/V_s \quad (48)$$

which can be expanded into (49) by substituting (47) therein:

$$p_{s4}^{\mathrm{ht}}(t)=m_sR(T_s-T_{RW})e^{-\frac{h_cA_c}{V_s\rho_{av}c_v}t}/V_s+m_sRT_{RW}/V_s \quad (49)$$

which can be reformed as

$$p_{s4}^{\mathrm{ht}}(t)=p_se^{-\frac{h_cA_c}{V_s\rho_{av}c_v}t}+m_sRT_{RW}\left(1-e^{-\frac{h_cA_c}{V_s\rho_{av}c_v}t}\right)/V_s \quad (50)$$

The $e^{-\frac{h_cA_c}{V_s\rho_{av}c_v}t}$ in (50) can be expressed as $e^{-\frac{h_cA_c}{m_{av}c_v}t}$, which can be linearized as follows

$$e^{-\frac{h_cA_c}{m_{av}c_v}t}=e^{-a_4}+\frac{a_4}{m_{av0}}e^{-a_4}(m_s-m_{av0}) \quad (51)$$

where $a_4=\frac{h_cA_ct}{m_{av0}c_v}$.

By substituting (51) into (47) and (50), one can obtain

$$T_{s4}^{\mathrm{ht}}(t)=(T_s-T_{RW})(e^{-a_4}+a_4e^{-a_4}(m_s-m_{av0})/m_{av0})+T_{RW} \quad (52)$$

$$p_{s4}^{\mathrm{ht}}(t)=p_s(e^{-a_4}+a_4e^{-a_4}(m_s-m_{av0})/m_{av0})+\frac{m_sRT_{RW}}{V_s}\left(1-e^{-a_4}-\frac{a_4}{m_{av0}}e^{-a_4}(m_s-m_{av0})\right) \quad (53)$$

There are three variables in (52), i.e., $T_{s4}^{\mathrm{ht}}(t)$, $T_s$, and $m_s$. There are also three variables in (53), i.e., $p_{s4}^{\mathrm{ht}}(t)$, $p_s$, and $m_s$.

In summary, the bi-linear cavern models include (34) and (38) for the charging process, (43) and (46) for the discharging process, and (52) and (53) for the idle process. Equations (34), (38), (43), (46), (52) and (53) are linear (bi-linear) equations when used in a one-step (multi-step) optimization problem.

# III. Simulation

Parameters for the Huntorf CAES plant are used for the calculations in this paper. The Huntorf CAES plant features two caverns with volumes of 141,000 and 169,000 m$^3$, respectively.

Note that the maximum mass flow rate in the charging process ($\dot{m}_{in}$) is 108 kg/s for the whole plant and 49.1226 kg/s for the first cavern, which is calculated from $108 \times \frac{141000}{141000+169000}$. Similarly, the maximum mass flow rate in the discharging process ($\dot{m}_{out}$) is 417 kg/s for the whole plant and 189.6677 kg/s for the first cavern, which is calculated from $417 \times \frac{141000}{141000+169000}$. In this paper, the first cavern is used for calculations. The other parameters for the Huntorf CAES plant are given in Table I [12], [13].

TABLE I
PARAMETERS FOR THE HUNTORF CAES PLANT.

| $A_c$ | $c_v$ | $h_c$ | $k$ | $p_{\text{in}}$ |
|---|---|---|---|---|
| 25,000 m$^2$ | 718.3 J/(kg K) | 30 W/(m$^2$ K) | 1.4 | 66 bar |
| $R$ | $V_s$ | $T_{RW}$ | $T_{\text{in}}$ | |
| 286.7 J/(kg K) | 141,000 m$^3$ | 40 °C | 50 °C | |

Three processes are defined as follows and used to verify the accuracy of the proposed bi-linear model in the rest of this section:

- Charging process: Set the initial pressure (temperature) of the air in the cavern to 46 bar (20 °C). Charge the first cavern (141,000 m$^3$) continuously for 16 hours at the maximum mass flow rate, i.e., $\dot{m}_{in}$ =49.1226 kg/s.
- Discharging process: Set the initial pressure (temperature) of the air in the cavern to 66 bar (40 °C). Discharge the cavern continuously for 4 hours at the maximum mass flow rate, i.e., $\dot{m}_{out}$ = 189.6677 kg/s.
- Idle process: Set the initial pressure (temperature) of the air in the cavern to 60 bar (45 °C). Let the cavern be in the idle process for 16 hours.

The status of the air in the cavern includes the temperature ($T_s$), pressure ($p_s$), and mass ($m_s$). When the status at the start of a time interval is known, which is used in the right-hand side of (34), (38), (43), (46), (52), and (53), then the status at the end of that time interval can be calculated, i.e., the left-hand side of (34), (38), (43), (46), (52), and (53), respectively.

The status of the air in the cavern in each interval of each process defined above is calculated in a recursive way. That is, given the status of the air at the start of a time interval, the status of the air at the end of that time interval is calculated via (34), (38), (43), (46), (52), and (53) as described in the previous paragraph, and is then used as the status of the air at the start of the next time interval.

Reference [13] compares several existing CAES models with the measured data from Huntorf. The analytical model in [13] is accurate and simpler than other existing analytical models. Thus, in the rest of this section, the analytical model in [13] is used as a benchmark model to verify the accuracy of the proposed bi-linear cavern model.

### *A. Model Verification*

In this section, the time interval is set to 1 second, i.e., $t$ is equal to 1 second in (34), (38), (43), (46), (52), and (53).

The pressure and temperature for each time interval of the charging (discharging, idle) process obtained from both the proposed bi-linear model and the analytical model given in [13] are plotted in Fig. 4 (Fig. 5, Fig. 6). Figs. 4-6 show that the pressure/temperature results obtained from both the proposed bi-linear model and the analytical model are quite close to one another.

The mean absolute percentage error between the results, in terms of pressure or temperature, obtained from the bi-linear model and the analytical model during the charging, discharging, and idle processes is tabulated in Table II. The last column of Table II shows that the idle part of the bi-linear model, i.e., (52) and (53), is almost as accurate as the analytical model. The 2$^{\text{nd}}$ and 3$^{\text{rd}}$ columns of Table II show that the accuracy of the charging/discharging parts of the bi-linear model, i.e., (34), (38), (43), and (46), is around 0.12%.

TABLE II
THE MEAN ABSOLUTE PERCENTAGE ERROR BETWEEN THE RESULTS OBTAINED BY THE BI-LINEAR MODEL AND THE ANALYTICAL MODEL GIVEN IN [13] IN EACH OF THE THREE PROCESSES.

| | Charging process | Discharging process | Idle process |
|---|---|---|---|
| Pressure | 0.0013 | 0.0012 | $1.61\times 10^{-7}$ |
| Temperature | 0.0012 | 0.0012 | $1.61\times 10^{-7}$ |

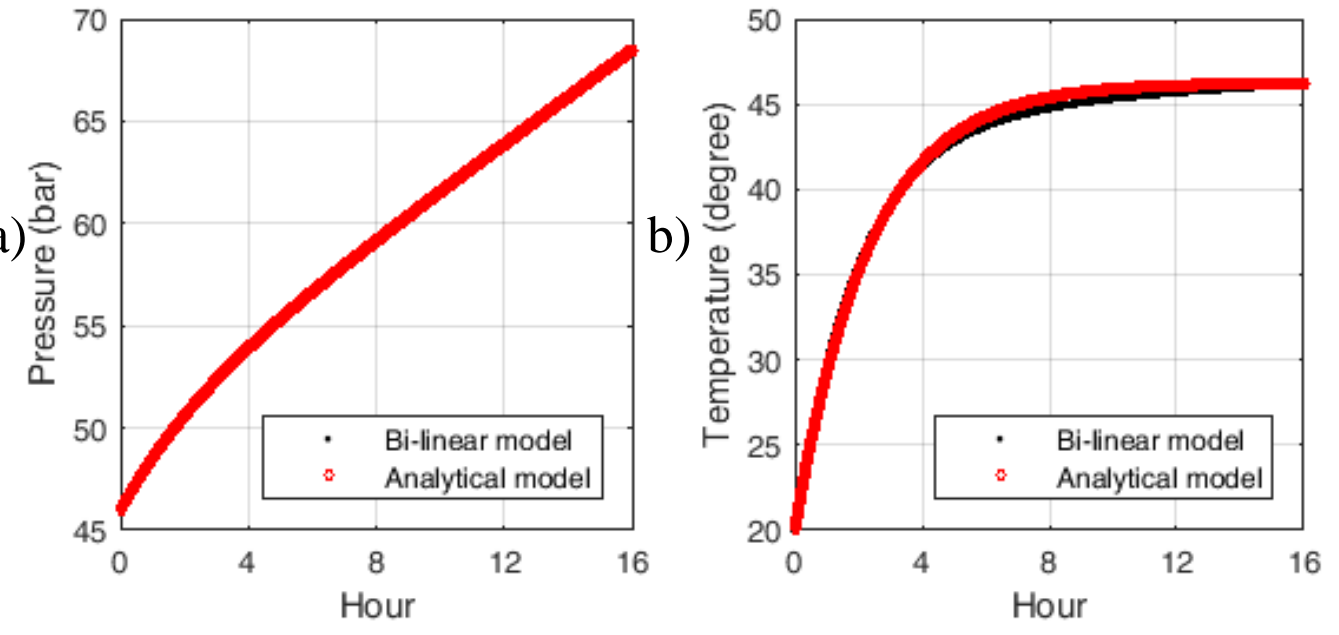

Fig. 4. Results obtained by the proposed bi-linear model and the analytical model in [13] during the charging process: a) pressure, b) temperature.

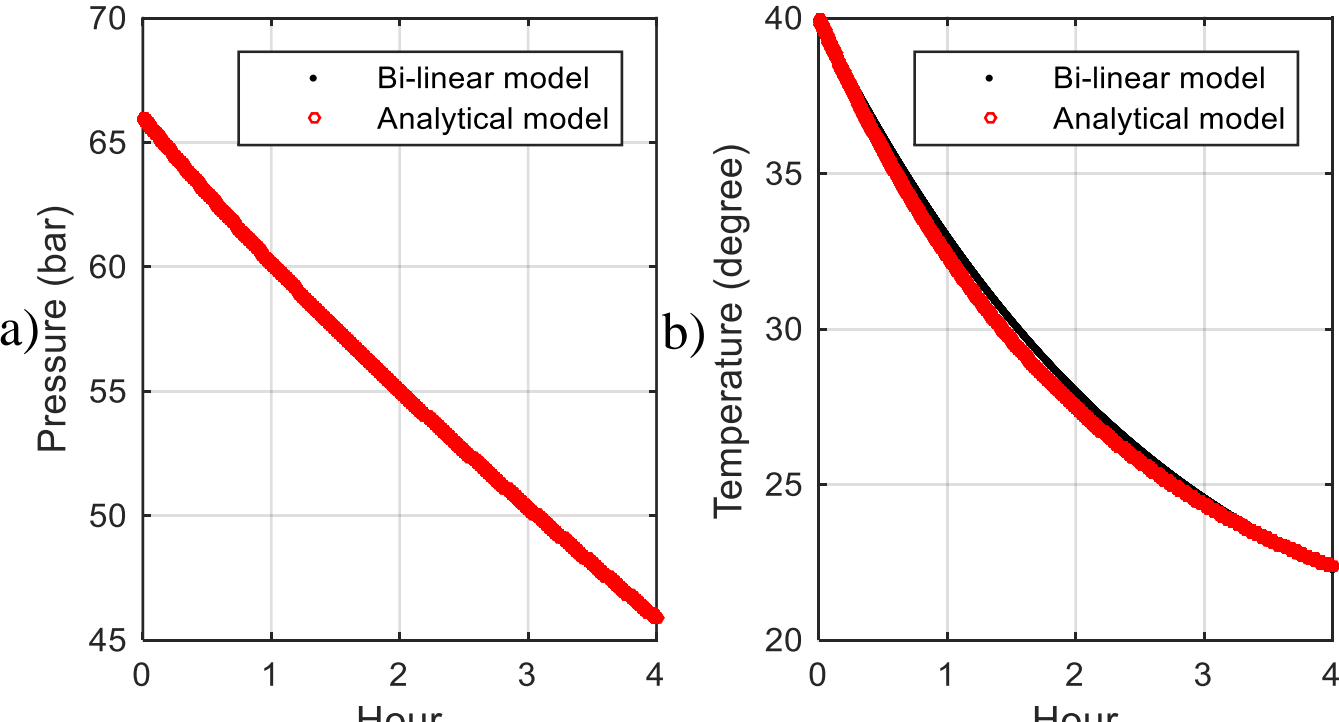

Fig. 5. Results obtained by the proposed bi-linear model and the analytical model in [13] during the discharging process: a) pressure, b) temperature.

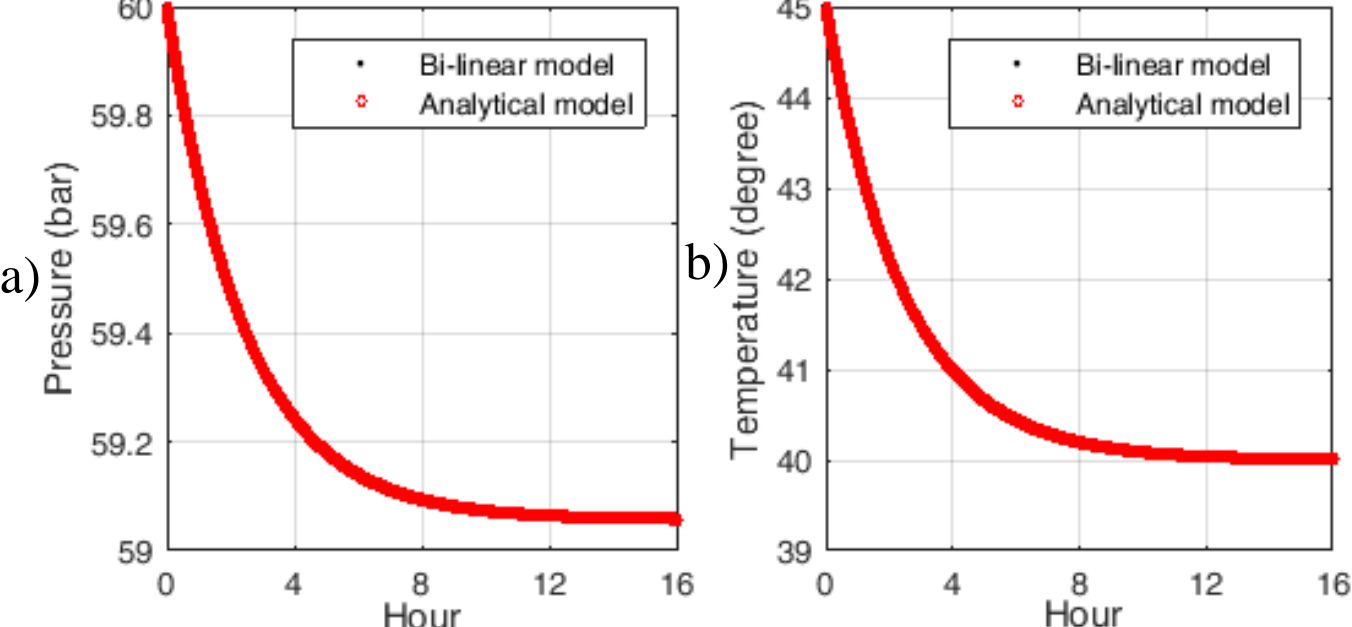

Fig. 6. Results obtained by the proposed bi-linear model and the analytical model in [13] during the idle process: a) pressure, b) temperature.

### B. *Impact of Heat Transfer and Temperature*

To observe the impact of the heat transfer, the results obtained from the bi-linear model with and without considering heat transfer in the charging (discharging) process are plotted in Fig. 7 (Fig. 8). The heat transfer clearly has a significant impact on the temperature and pressure. Therefore, it is important to consider the heat transfer in the cavern model.

The results obtained from the constant temperature model in the charging (discharging) process are also plotted in Fig. 7 (Fig. 8). Obviously, the pressure and temperature obtained from the constant temperature model are quite different from the bi-linear model, which indicates that the constant temperature cavern model is inaccurate. Therefore, it is necessary to use an accurate cavern model instead of the constant temperature cavern model.

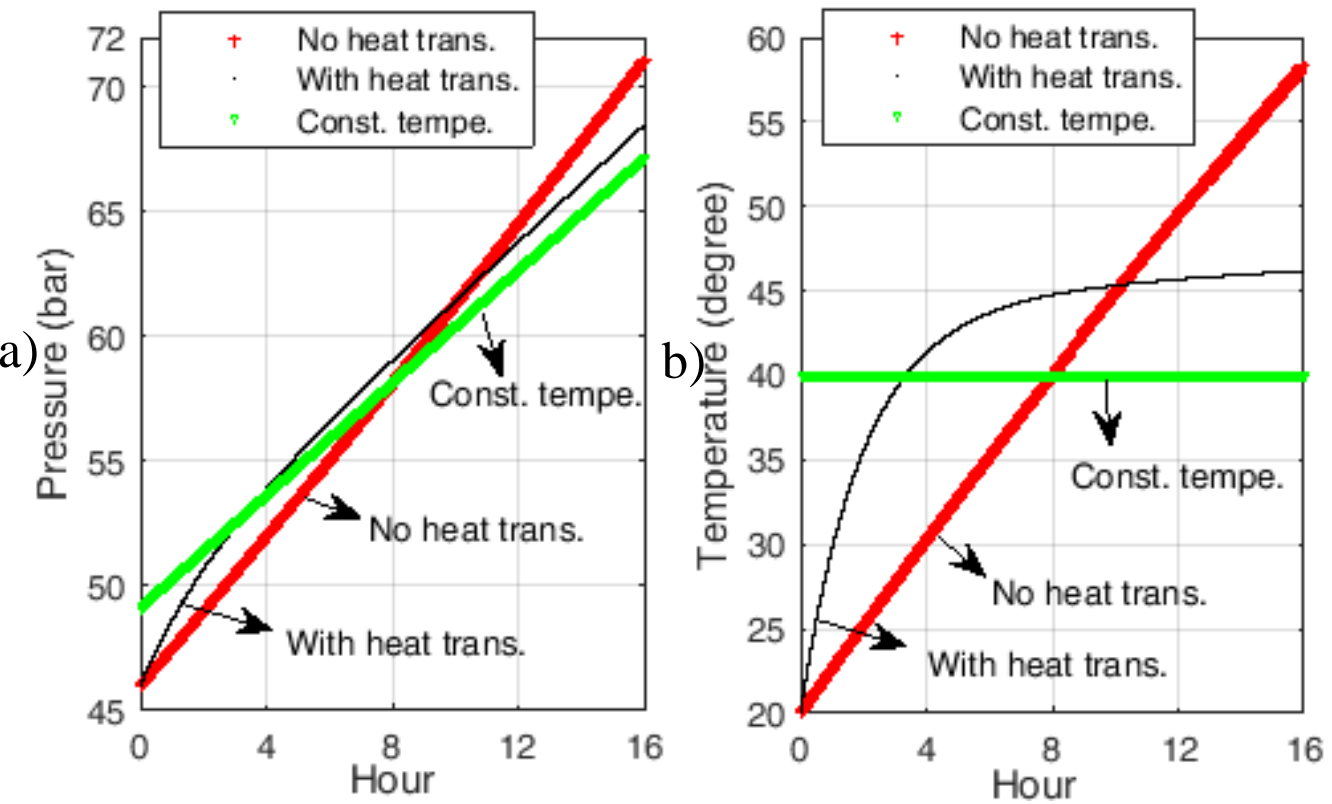

Fig. 7. Results obtained from the bi-linear cavern model with and without considering heat transfer and the constant temperature cavern model in the charging process: a) pressure of the air in the cavern, and b) temperature of the air in the cavern.

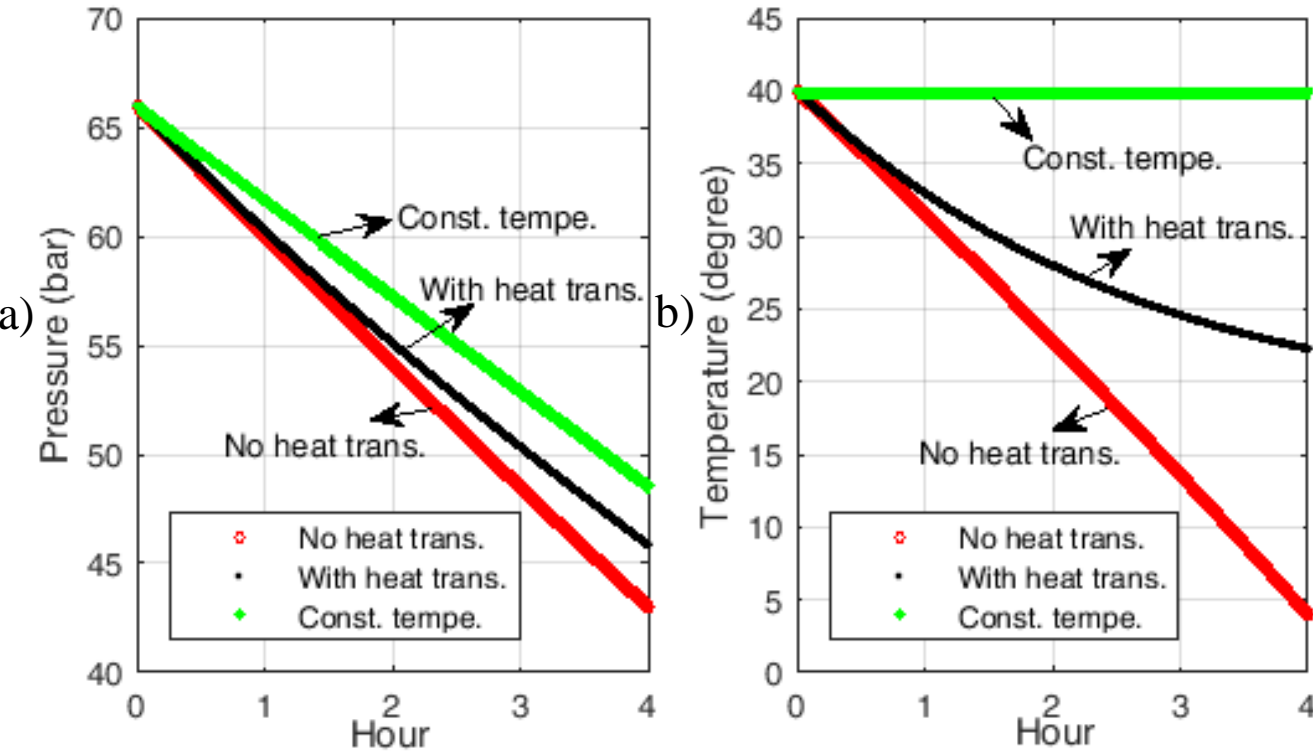

Fig. 8. Results obtained from the bi-linear cavern model with and without considering heat transfer and the constant temperature cavern model in the discharging process: a) pressure of the air in the cavern, and b) temperature of the air in the cavern.

### C. *Different Time Intervals*

In power system operation problems, the time interval can be longer than one second, e.g., the time intervals of economic dispatch and unit commitment are usually 15 minutes and 1 hour, respectively. Therefore, it is necessary to know whether the proposed bi-linear model is accurate for different time intervals. In this regard, the status of the air in the cavern is calculated for different time intervals (1 second, 1 minute, 5 minutes, 10 minutes, 20 minutes, and 60 minutes) using the same initial status and the same recursive approach as outlined in the second paragraph before Section III-A. The final temperature and pressure of the charging, discharging, and idle processes obtained by the bi-linear model and the analytical model are plotted in Fig. 9. The error and relative error (values given in round brackets) between the results, in terms of final temperature and pressure, obtained from the two models are shown in Table III, where the $2^{nd}$-$5^{th}$, the $6^{th}$-$9^{th}$, and the $10^{th}$-$13^{th}$ rows show the error/relative error in the charging, discharging, and idle processes, respectively. In Table III, 'E-4' and 'E-6' represent '$\times 10^{-4}$' and '$\times 10^{-6}$', respectively.

Figs. 9a-9d show that the accuracy of the bi-linear model in the charging and discharging processes decreases as the time interval increases. Figs. 9e-9f show that the accuracy of the bi-linear model in the idle process does not change with the time interval.

Table III shows that the error and relative error of the temperature and pressure in both the charging and discharging processes are small when the time interval is smaller than or equal to 5 minutes. When the time interval is equal to 10 or 20 minutes, the errors (relative errors) of the pressure in the charging process are 0.0416 and 0.0993 bar (0.06 and 0.14%), respectively, which are relatively small. When the time interval is equal to 10 or 20 minutes, the errors (relative errors) of the pressure in the discharging process are 0.0298 and 0.0740 bar (0.06 and 0.16%), respectively, which are also relatively small. When the time interval is equal to 60 minutes, the error (relative error) in both the charging and discharging processes is relatively large. Therefore, Fig. 9 and Table III show that the accuracy of the bi-linear model is high, moderate, and relatively low when the time interval is between 1 second and 5 minutes, between 10 and 20 minutes, and equal to 60 minutes, respectively.

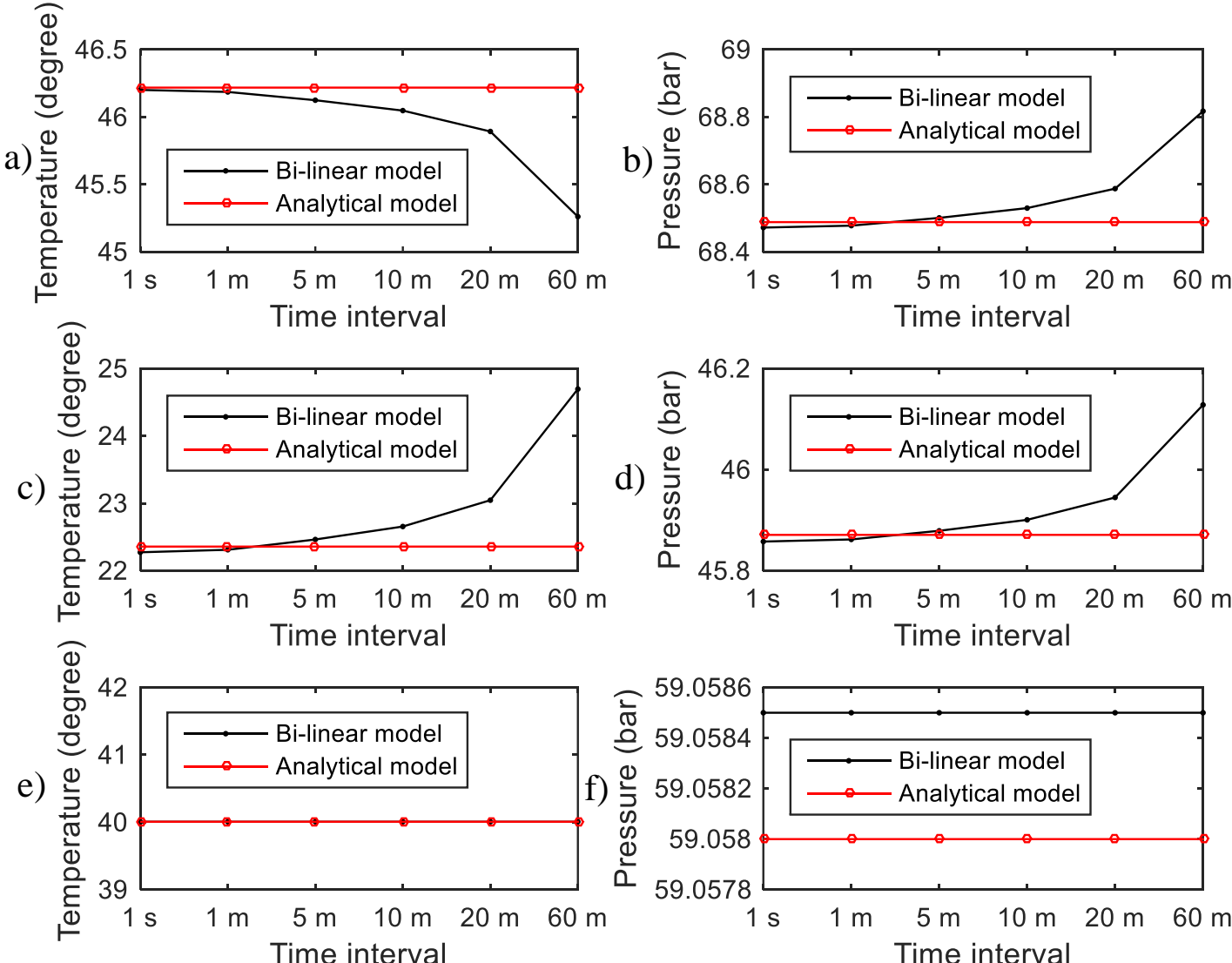

Fig. 9. Final temperature/pressure obtained by both the analytical model and the bi-linear model in different processes using different time intervals: a) final temperature in the charging process, b) final pressure in the charging process, c) final temperature in the discharging process, d) final pressure in the discharging process, e) final temperature in the idle process, f) final pressure in the idle process.

TABLE III
ERROR (RELATIVE ERROR IN ROUND BRACKETS) BETWEEN THE SOLUTION OBTAINED BY THE BI-LINEAR MODEL AND THE ANALYTICAL MODEL IN DIFFERENT PROCESSES USING DIFFERENT TIME INTERVALS (2ND-5TH ROWS FOR CHARGING PROCESS, 6TH-9TH ROWS FOR DISCHARGING PROCESS, 10TH-13TH ROWS FOR IDLE PROCESS)

| Interval | 1 s | 1 min | 5 min | 10 min | 20 min | 60 min |
|---|---|---|---|---|---|---|
| Temperature (°C) | -0.0173 (-0.04%) | -0.0323 (-0.07%) | -0.0936 (-0.2%) | -0.1706 (-0.37%) | -0.3255 (-0.7%) | -0.9574 (-2.1%) |
| Pressure (bar) | -0.0162 (-0.02%) | -0.0105 (-0.02%) | 0.0126 (0.02%) | 0.0416 (0.06%) | 0.0993 (0.14%) | 0.3292 (0.48%) |
| Temperature (°C) | -0.0848 (-0.38%) | -0.0475 (-0.21%) | 0.1050 (0.47%) | 0.2973 (1.3%) | 0.6879 (3.1%) | 2.3371 (10%) |
| Pressure (bar) | -0.0132 (-0.03%) | -0.0090 (-0.02%) | 0.0081 (0.02%) | 0.0298 (0.06%) | 0.0740 (0.16%) | 0.2572 (0.56%) |
| Temperature (°C) | 0 (0) | 0 (0) | 0 (0) | 0 (0) | 0 (0) | 0 (0) |
| Pressure (bar) | 5E-4 (8.5E-6) | 5E-4 (8.5E-6) | 5E-4 (8.5E-6) | 5E-4 (8.5E-6) | 5E-4 (8.5E-6) | 5E-4 (8.5E-6) |

## IV. CONCLUSION

This paper has proposed an accurate bi-linear cavern model for CAES based on the ideal gas law and the first law of thermodynamics. An accurate analytical model in the literature is used as benchmark model to verify the accuracy of the proposed bi-linear cavern model.

Simulation results show that the error between the bi-linear model and the accurate analytical model is around 0.12% when the time interval is set to 1 second. The accuracy of the proposed bi-linear model decreases as the time interval increases. For time intervals between 1 second and 5 minutes, between 10 and 20 minutes, and 60 minutes or longer, the bi-linear cavern model has high, moderate, and relatively low accuracy, respectively. Simulation results also show that heat transfer has an obvious effect on the variation of temperature and pressure of the air in the cavern. Therefore, it is necessary to consider heat transfer in the cavern model. The constant-temperature cavern model is also shown to be inaccurate, which emphasizes the necessity of the proposed bi-linear cavern model for power system optimization problems.

By properly setting the time interval, the proposed bi-linear cavern model is accurate and suitable for use in power system optimization problems, as will be demonstrated in the second paper of this two-part series.

**Junpeng Zhan** (M'16) received B.Eng. and Ph.D. degrees in electrical engineering from Zhejiang University, Hangzhou, China in 2009 and 2014, respectively.

He was a Postdoctoral Fellow in the Department of Electrical and Computer Engineering, University of Saskatchewan, Saskatoon, SK, Canada. His current research interests include the integration of the energy storage systems, dynamic thermal rating and renewable electric energy sources in power systems.

**Osama Aslam Ansari** (S'16) received the B.Eng. degree in electrical engineering from National University of Sciences and Technology (NUST), Islamabad, Pakistan, in 2015.

He is currently working toward the M.Sc. degree in electrical engineering at the Department of Electrical and Computer Engineering, University of Saskatchewan, Saskatoon, SK, Canada. His current research interests include the energy storage systems and power system reliability.

**C. Y. Chung** (M'01-SM'07-F'16) received B.Eng. (with First Class Honors) and Ph.D. degrees in electrical engineering from The Hong Kong Polytechnic University, Hong Kong, China, in 1995 and 1999, respectively.

He has worked for Powertech Labs, Inc., Surrey, BC, Canada; the University of Alberta, Edmonton, AB, Canada; and The Hong Kong Polytechnic University, China. He is currently a Professor, the NSERC/SaskPower (Senior) Industrial Research Chair in Smart Grid Technologies, and the SaskPower Chair in Power Systems Engineering in the Department of Electrical and Computer Engineering at the University of Saskatchewan, Saskatoon, SK, Canada. His research interests include smart grid technologies, renewable energy, power system stability/control, planning and operation, computational intelligence applications, and power markets.

Dr. Chung is an Editor of *IEEE Transactions on Power Systems*, *IEEE Transactions on Sustainable Energy*, and *IEEE Power Engineering Letters* and an Associate Editor of *IET Generation, Transmission, and Distribution*. He is also an IEEE PES Distinguished Lecturer and a Member-at-Large (Global Outreach) of the IEEE PES Governing Board.